\documentclass{article}
\usepackage[dblblindworkshop, final]{neurips_2026}
\usepackage[utf8]{inputenc}
\usepackage[T1]{fontenc}
\usepackage{hyperref}
\usepackage{url}
\usepackage{booktabs}
\usepackage{amsfonts}
\usepackage{amsmath}
\usepackage{amssymb}
\usepackage{nicefrac}
\usepackage{microtype}
\usepackage{graphicx}
\usepackage{xcolor}
\usepackage{subcaption}
\usepackage{multirow}
\graphicspath{{figures/}}

\newcommand{\green}[1]{\textcolor{green!55!black}{#1}}
\newcommand{\red}[1]{\textcolor{red!70!black}{#1}}
\newcommand{\up}[1]{\textcolor{green!55!black}{$\uparrow$#1}}
\newcommand{\dn}[1]{\textcolor{red!70!black}{$\downarrow$#1}}
\newcommand{\sid}{\textsc{sid}}

\title{Personalized LLM Recommendation with Semantic IDs: Autoregressive Beam Search vs. Multi-Token Prediction at Industry Scale}
\workshoptitle{Personalized, aligned, long-term memory for AI systems}

\author{%
  Sejoon Oh\thanks{Project lead / corresponding author: \texttt{sejoono@netflix.com}.} \quad
  Fernando Amat Gil \quad Evan Smothers \quad Dan Iter \quad Mark Thornburg \\
  \textbf{Dawit Mureja Argaw} \quad
  \textbf{Sourabh Medapati} \quad \textbf{Thea Wang} \quad 
  \textbf{Ria Verma} \quad \textbf{Qiuling Xu}
  \\
  \textbf{Harshit Syal} \quad \textbf{Ehsan Gholami} \quad \textbf{Moumita Bhattacharya} \quad
  \textbf{Ashish Rastogi} \\
  \\
  Netflix Personalization Foundations Team \\
}

\begin{document}

\maketitle

\begin{abstract}
Large-scale personalization is shifting from embedding-table retrieval to
LLM-based recommenders that \emph{generate} the \emph{Semantic ID} (\sid{}) of
the item to recommend, where each \sid{} is a short sequence of discrete hierarchical content
codes. A single instruction-tuned model can then unify retrieval, content
understanding, and language, and recommend brand-new titles from their content
alone using \sid{}; then, \emph{how} should the model generate the \sid{}? Traditionally, models rely on a single language modeling head that generates tokens autoregressively, typically using constrained beam search. In contrast, an emerging body of work explores multi-token prediction (MTP), where dedicated heads predict all tokens in parallel, substantially reducing the decoding latency associated with beam search. We report an
\emph{industry-scale, apples-to-apples} comparison of these two \emph{generation
paradigms}: single-head autoregressive decoding vs.\ MTP.
We introduce a new auxiliary ranking loss, defined over the joint SID score, to train the MTP objective. 
We evaluate the \sid{} training objectives on both quality and training/inference cost, inside a multi-task LLM that jointly does personalized retrieval, content
question-answering (QA), and instruction following over a large streaming-media catalog. Our experiments show \textbf{single-head beam search consistently beats
MTP on most product-critical metrics}: ranking and content QA, and trains at roughly
\textbf{2.6x} lower dollar cost on \textit{Qwen3-8B} scale; MTP wins only on raw inference
throughput. Around this finding, we detail an efficient training and inference
stack and report end-to-end gains over the \emph{previous non-\sid{} baseline equipped with an auxiliary ranking head}: \textbf{+77\%}
item cold-start ranking, \textbf{+22\%} content-QA accuracy, \textbf{+2.8x} faster training, with language metric regression traced to \sid{} vocabulary expansion. We address the language regression by reinforcement learning post training.
\end{abstract}

\section{Introduction}
\label{sec:intro}

Recent large-scale personalization systems increasingly leverage a large language model (LLM) to unify retrieval, ranking, content understanding, and natural-language understanding. Such an LLM ingests a member's interaction
history and context and must (i) \emph{retrieve} relevant catalog entities,
(ii) \emph{understand} title content well enough to \emph{answer}
natural-language questions about it (content QA), and (iii) \emph{follow} general
instructions. We treat content understanding as a first-class training and evaluation objective alongside ranking, since it lets the model
reason about \emph{why} a title fits a member, ground recommendations in metadata (not just popularity), and generalize to titles with little or no engagement history. 

Modern systems refer to catalog entities not by arbitrary integer ids but by \emph{Semantic IDs} (\sid{}s): short sequences of discrete codes produced by residual quantization
\citep{lee2022rqvae, tay2022dsi,rajput2023tiger,singh2024semanticid} of a content embedding, so that semantically similar titles
share early-level codes.
The model then \emph{generates} the target's \sid{} on its own output head, which removes the catalog-sized classification head, unifies retrieval with the language and content-understanding tasks, and greatly improves
cold-start because a new title inherits a
meaningful (non-random) \sid{} from its content embedding. This recipe now underlies several industrial systems (YouTube's PLUM
\citep{he2025plum}, Spotify's GLIDE \citep{damico2026glide}, Kuaishou's OneRec
\citep{deng2025onerec}, Meta's generative recommenders \citep{zhai2024hstu}),
which report substantial online gains.

This raises the question at the focus of this paper: \emph{how should the LLM
generate \sid{}?} \textbf{(a) Single-head autoregressive} decoding emits the
codes one at a time on the standard LM head with constrained beam search, which is also the
classical recipe of DSI and TIGER \citep{tay2022dsi,rajput2023tiger} and the
fine-tuning stage of PLUM \citep{he2025plum} and GLIDE \citep{damico2026glide}.
\textbf{(b) Multi-token prediction (MTP)} attaches $L$ dedicated heads to a
shared trunk and predicts all codes in parallel \citep{gloeckle2024mtp}. 
Leveraging this parallel decoding, MTP can generate \sid{} faster than the beam search in inference time. 
Only a few studies such as  RPG \citep{hou2025rpg}
and STAMP \citep{zhan2026semantic} have applied MTP in generative recommendations. Those studies achieve higher ranking metrics or training efficiency by leveraging parallel \sid{} prediction.

However, these arguments are often made on small, retrieval-only academic benchmarks. Using an open-source model with an industry-scale streaming dataset as a controlled testbed, we reach the conclusion in the regime that
matters for deployment: \textbf{single-head constrained beam search beats MTP on
most product-critical metrics} (ranking and content QA), with a significantly lower training cost on \textit{Qwen3-8B} scale. Meanwhile, MTP outperforms on inference throughput only.

Our contributions are:
\begin{itemize}
  \item \textbf{The industry-scale, apples-to-apples comparison of
  single-head autoregressive decoding vs.\ MTP as \sid{} generation paradigms.} 
  A \emph{quality-and-cost} comparison of training objectives, inside a unified
  multi-task LLM and large-scale streaming dataset, on ranking, content QA, language, and
  training/inference performance (Section~\ref{sec:core}). 
  \item \textbf{An auxiliary ranking loss for MTP}, a new objective that
  optimizes the item-level joint \sid{} score so that MTP's parallel heads produce calibrated, rankable item scores
  despite their conditional-independence assumption
  (Section~\ref{sec:method-mtp}). 
  \item \textbf{An efficient training and inference stack}: Muon
  \citep{jordan2024muon}, on-the-fly packing with cross-rank batch mixing, Cut
  Cross-Entropy \citep{wijmans2024cce}, and HSDP \citep{zhao2023fsdp} for
  training ($>2\times$ faster at $-73\%$ cost); and \textbf{vLLM-based
  constrained, catalog-aware beam-search decoding} for inference
  \citep{kwon2023vllm}, whose trie- and catalog-level constraints guarantee every
  candidate names a real, request-eligible entity while prefix caching reuses the
  long member-history KV cache across beams (Sections~\ref{sec:method-inf}--\ref{sec:method-stack}).
  \item \textbf{A comprehensive evaluation and ablations} covering new user and item recommendations, plus
  controlled ablations of LoRA~\citep{hu2022lora} vs.\ full fine-tuning, Muon~\citep{jordan2024muon} vs. AdamW~\citep{loshchilov2018decoupled} optimizers, and content QA tasks
  (Section~\ref{sec:experiments}).
\end{itemize}

\section{Related Work}
\label{sec:related}

\paragraph{Generative retrieval and semantic IDs.}
The Differentiable Search Index \citep{tay2022dsi} first framed retrieval as
sequence-to-sequence generation of document identifiers. TIGER
\citep{rajput2023tiger} introduced \emph{semantic IDs} for recommendation,
obtaining discrete codes from an RQ-VAE \citep{lee2022rqvae} over item content
embeddings and autoregressively generating the next item's id.
\citep{singh2024semanticid} show semantic IDs generalize better than random
atomic ids for ranking. At industrial scale, PLUM \citep{he2025plum} adapts a
pre-trained LLM to generative retrieval via continued pre-training on
\sid{}-tokenized behavior, GLIDE \citep{damico2026glide} frames podcast discovery
as instruction-following over a \sid{}-discretized catalog, and OneRec
\citep{deng2025onerec} unifies retrieval and ranking in one generative model. We
adopt this representation but place the \sid{} vocabulary on a general-purpose
instruction-tuned LLM's own head, so that retrieval, content QA, and language
tasks share a single decoder, rather than training a bespoke retrieval-only
transformer.

\paragraph{Autoregressive beam search vs.\ parallel/MTP decoding.}
MTP was introduced as an auxiliary objective that predicts
$n$ future tokens with $n$ independent heads on a shared trunk, improving sample
efficiency and enabling self-speculative decoding \citep{gloeckle2024mtp}; it is
now a component of frontier LLMs \citep{deepseek2024v3}. In generative
recommendation (GR), a recent wave applies this idea to \sid{}s to escape the
autoregressive bottleneck. RPG \citep{hou2025rpg} predicts long (up to 64-token)
\emph{unordered} \sid{}s in parallel with an MTP loss and graph-constrained
decoding for higher efficiency and ranking quality; STAMP \citep{zhan2026semantic} introduces an auxiliary multi-token prediction objective to densify supervision, improving representation learning while semantic pruning reduces training cost.
\citep{wu2025constrained} question constrained decoding's generalization
theoretically. We revisit these conclusions \emph{empirically and at industry
scale} (Section~\ref{sec:core}): within a unified multi-task LLM, and
even after strengthening MTP with an auxiliary ranking loss, single-head beam
search delivers better ranking and content-QA quality at lower training cost,
while MTP's advantage is limited to inference throughput and full-catalog ranking capability\footnote{It is computationally infeasible for the beam search to generate probabilities across entire catalog.}.

\paragraph{Speculative and self-speculative decoding.}
A distinct line of work uses multi-token or draft heads purely as a
\emph{lossless accelerator}: Medusa \citep{cai2024medusa} adds parallel decoding
heads verified by tree attention, Speculative Streaming
\citep{bhendawade2024specstream} folds future $n$-gram drafting into the target
model, and \citep{zagyva2025medusakd} combine Medusa with distillation for an
industrial travel application. Crucially, these methods speed up inference while
\emph{preserving} the underlying autoregressive model's output distribution, so
they do not trade quality for speed; they are orthogonal to, and can be layered
on top of, our single-head decoder rather than being an alternative to it. NEZHA ~\citep{wang2026nezha} makes the
distinction explicit: it accelerates \sid{} beam search with an
\emph{autoregressive} self-drafting head plus a hash-set verifier, and reports
that a plain MTP decoder is faster but markedly worse in quality because it
cannot exploit the sequential dependency inherent in the \sid{}. We instead compare single-head vs.\ MTP as
\emph{generation paradigms} (a quality-and-cost question), whereas this line
targets quality-preserving acceleration (a speed question).

\paragraph{Efficient LLM training and inference.}
Our stack combines existing components: Muon \citep{jordan2024muon}, Cut
Cross-Entropy \citep{wijmans2024cce}, sequence packing \citep{krell2023efficient},
and HSDP \citep{zhao2023fsdp} for both single-head and MTP training, and vLLM's PagedAttention and prefix
caching \citep{kwon2023vllm} for inference with the single-head and beam search. Our contribution here is not any single
component but their integration and measured joint effect on a large-scale run, plus the trie- and catalog-constrained decoding layer we build on vLLM.

\section{Method}
\label{sec:method}

We first define how entities are represented and generated as Semantic IDs
(Section~\ref{sec:method-sid}), then the two \emph{training} paradigms we compare: single-head and MTP, the latter equipped with our auxiliary ranking loss
(Section~\ref{sec:method-train}). We also introduce their corresponding \emph{inference}
procedures (Section~\ref{sec:method-inf}), and finally the efficient
training/inference stack (Section~\ref{sec:method-stack}).

\subsection{Semantic-ID Generation}
\label{sec:method-sid}

Let $\mathcal{E}$ be the catalog of entities. Each entity $e\in\mathcal{E}$ is
mapped to a \emph{Semantic ID}
\begin{equation}
  \mathrm{SID}(e) = \big(c_1(e), c_2(e), \dots, c_L(e)\big), \qquad
  c_\ell(e)\in\{1,\dots,K_\ell\},
  \label{eq:sid}
\end{equation}
a length-$L$ sequence of discrete codes drawn from per-level codebooks of sizes
$K_1,\dots,K_L$, obtained by residual quantization \citep{lee2022rqvae} of a
content embedding $\phi(e)\in\mathbb{R}^d$. The quantizer is trained so that the
first code captures the coarsest semantic partition and later codes refine it, so
entities close in content space share early-level codes. The framework is
agnostic to the \sid{} \emph{shape}: level sizes may be uniform (e.g.,
$1024\times4$, meaning $K_\ell{=}1024$ for all $L{=}4$ levels) or pyramidal
(e.g., $1024{\times}512{\times}256{\times}128$, spending more capacity on the
coarse levels); our experiments use the uniform $1024\times4$ shape. The
$\sum_\ell K_\ell$ \sid{} tokens are appended to the LLM vocabulary $\mathcal{V}$,
giving $\mathcal{V}^{+}=\mathcal{V}\cup\mathcal{S}$ with
$|\mathcal{S}|=\sum_{\ell=1}^{L} K_\ell$. While \sid{} collisions rarely happen (e.g., $\leq 1\%$), we resolve the collision by returning the first entity in the \sid{} bucket.

We place the \sid{} vocabulary on the LM head and \emph{generate}
the target's codes, rather than scoring a catalog-sized auxiliary head as the
baseline did. This gives (i) a \emph{simpler architecture} that has one model, one
head, with no catalog-sized layer to provision or re-train as the catalog grows;
(ii) \emph{stronger cold-start}, since a new title's \sid{} is computed from its
content embedding and can be recommended before any member interacts with it, by
sharing \sid{} prefixes with established titles\footnote{For ``extreme''
cold-start entities whose \sid{}s are never exposed during training, we fall back
to retrieval-augmented prompting or lightweight daily fine-tuning.}, and (iii)
\emph{shared representations}, since retrieval, content QA, and instruction
following reuse the same head.

\subsection{Training}
\label{sec:method-train}

\begin{figure}[t]
  \centering
  \includegraphics[width=1.0\linewidth]{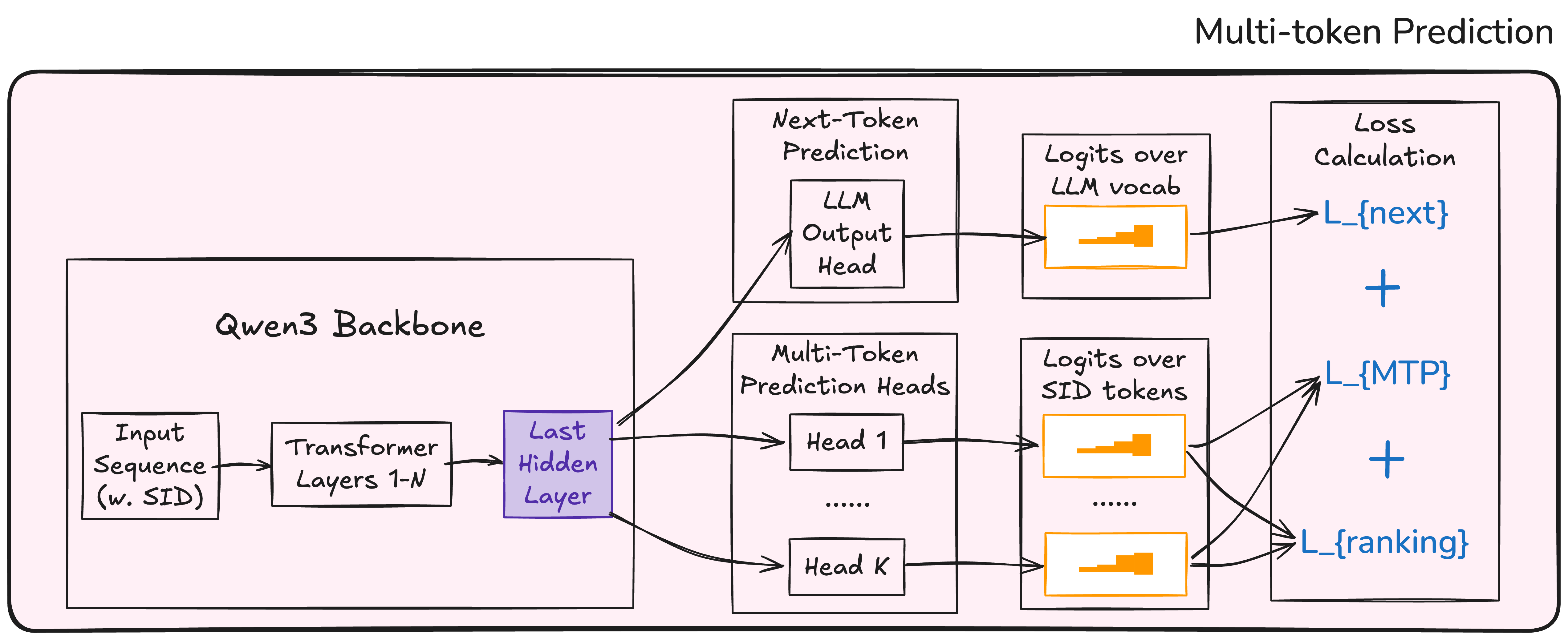}
  \caption{\textbf{Multi-token-prediction (MTP) training.} While the single-head approach uses a LM head with a next-token loss
  (Eq.~\ref{eq:loss}), MTP adds $L$ parallel \sid{} heads trained with per-code
  cross-entropy \emph{and} our auxiliary ranking loss over the joint \sid{} score
  (Eq.~\ref{eq:mtp}).}
  \label{fig:mtp-arch}
\end{figure}

Every training example (e.g., retrieval, content QA, or instruction following) is
a token sequence over $\mathcal{V}^{+}$: retrieval targets are written as the
target's \sid{} tokens, content QA references titles by both text and \sid{}, and
instruction data is unchanged. We compare two ways of producing the \sid{} codes;
both share the same backbone, tokenizer, and data (Figure~\ref{fig:mtp-arch}).

\subsubsection{Single-Head (LM Head) Training}
\label{sec:method-single}
The classical design uses the single next-token LM head to emit \sid{} codes
\emph{autoregressively}, so training is one next-token problem whose per-example
loss splits into text and \sid{} positions,
\begin{equation}
  \mathcal{L}^{\text{SH}}
  = -\!\!\sum_{i\in \mathcal{T}_{\text{text}}}\!\! \log p_\theta(y_i\mid y_{<i})
  \;-\; \lambda \!\!\sum_{i\in \mathcal{T}_{\text{sid}}}\!\! \log p_\theta(y_i\mid y_{<i}),
  \label{eq:loss}
\end{equation}
where $\mathcal{T}_{\text{text}}$ and $\mathcal{T}_{\text{sid}}$ index text and
\sid{} target positions and $\lambda\ge 1$ up-weights the sparser \sid{}
positions. Crucially, this autoregressive training enforces dependencies between the SID levels, where predicting the current code $c_\ell$ is conditioned on previous codes $c_{<\ell}$. During inference, beam search retains multiple early prefixes that later scores can rerank.

\subsubsection{Multi-Token Prediction (MTP) Training}
\label{sec:method-mtp}
Following \citep{gloeckle2024mtp}, MTP keeps the next-token LM head and adds $L$
parallel heads $g_1,\dots,g_L$ on the shared last-hidden state $h_x$ for query
$x$, with head $g_\ell$ responsible for code $c_\ell$. Because all codes are
predicted \emph{in parallel} from $h_x$, the standard MTP retrieval term is a sum
of independent per-code cross-entropies,
\begin{equation}
  \mathcal{L}_{\text{code}}
  \;=\; -\sum_{\ell=1}^{L}\log p_{g_\ell}\!\big(c_\ell(e^{+}) \mid h_x\big),
  \label{eq:mtpcode}
\end{equation}
for the positive item $e^{+}$. This factorization is exactly the weakness of MTP
for ranking: it optimizes each code's \emph{marginal} likelihood but never the
\emph{joint} item score, so an off-target item whose individual codes are each
independently probable can outscore the true target.

\paragraph{Auxiliary ranking loss.}
We address this with a new auxiliary objective that directly optimizes the joint,
item-level score. Define the MTP score of a candidate $e$ as the sum of its
per-code log-probabilities,
\begin{equation}
  s_{\text{MTP}}(e\mid x) \;=\; \sum_{\ell=1}^{L}\log p_{g_\ell}\!\big(c_\ell(e)\mid h_x\big),
  \label{eq:mtpscore}
\end{equation}
and train the heads so the positive item ranks above a set $\mathcal{N}(x)$ of
sampled in-batch/hard negatives from the eligible catalog, via a
sampled-softmax \footnote{If the entire catalog is small (e.g., $\leq 1M$), normal Softmax without any sampled negatives can be used here.}  ranking loss:
\begin{equation}
  \mathcal{L}_{\text{rank}}
  \;=\; -\log
  \frac{\exp\!\big(s_{\text{MTP}}(e^{+}\mid x))}
       {\displaystyle\sum_{e\in\{e^{+}\}\cup\mathcal{N}(x)}\exp\!\big(s_{\text{MTP}}(e\mid x))},
  \qquad
  \mathcal{L}^{\text{MTP}}
  = \mathcal{L}_{\text{next}} + \alpha\,\mathcal{L}_{\text{code}}
  + \beta\,\mathcal{L}_{\text{rank}},
  \label{eq:mtp}
\end{equation}
with importance hyperparameters $\alpha,\beta$, and $\mathcal{L}_{\text{next}}$
the text next-token loss retained from the LM head. Unlike standard MTP
\citep{gloeckle2024mtp}, which uses per-token cross-entropy only, and RPG
\citep{hou2025rpg}, which pairs MTP with graph-constrained decoding at inference,
$\mathcal{L}_{\text{rank}}$ teaches the \emph{independent} heads to produce
jointly calibrated, rankable item scores at \emph{training} time, which partially
compensates for the missing autoregressive coupling and gives MTP its strongest
form. We empirically confirm that having $\mathcal{L}_{\text{rank}}$ improves ranking metrics, while increasing training cost.

\subsection{Inference}
\label{sec:method-inf}

\subsubsection{Single-Head: Constrained, Catalog-Aware Beam Search}
\label{sec:method-beam}
A single-head retrieval call generates the target \sid{} token-by-token. Let
$\mathcal{T}$ be the \emph{\sid{} trie} of all prefixes realized by some entity,
and $\mathcal{C}_x\subseteq\mathcal{E}$ the per-request eligible catalog for
query $x$. At step $\ell$ with prefix $c_{<\ell}$, we restrict the next-token
distribution to the admissible set
\begin{equation}
  \mathcal{A}(c_{<\ell}, x) \;=\;
  \big\{\, k\;:\; (c_{<\ell},k)\in \mathcal{T}
  \ \text{and}\ \exists\, e\in\mathcal{C}_x,\ \mathrm{SID}(e)\!\succ\!(c_{<\ell},k)\,\big\},
  \label{eq:admissible}
\end{equation}
where $\succ$ denotes ``has as a prefix''. \emph{Constrained} decoding
guarantees we never emit a \sid{} that names no real entity; \emph{catalog-aware}
decoding guarantees every candidate is eligible for the request, sharpening local
ranking metrics without post-hoc filtering. Beam search keeps the top-$N$ partial
hypotheses per level, scoring a complete \sid{} by its length-normalized
log-probability
\begin{equation}
  s(e\mid x) \;=\; \frac{1}{L}\sum_{\ell=1}^{L}
  \log p_\theta\!\big(c_\ell(e) \mid x, c_{<\ell}(e)\big),
  \label{eq:beamscore}
\end{equation}
and returns the top-$K$ entities. We implement this on vLLM~\citep{kwon2023vllm} that reuses the long member-history prompt's KV
cache across beams and decoding steps, so wide beams are affordable, and the
trie/catalog constraints are applied as a logit mask at each step.

\subsubsection{MTP: Parallel Decoding}
\label{sec:method-mtpinf}
MTP produces all $L$ per-code distributions from $h_x$ in a \emph{single} forward
pass and scores candidates by $s_{\text{MTP}}$ (Eq.~\ref{eq:mtpscore}); the
top-$K$ eligible entities are selected by evaluating $s_{\text{MTP}}$ over the
trie-restricted candidate set, without sequential beam expansion. This removes
the per-code forward passes of beam search, which is the source of MTP's inference-throughput
advantage. However, it also removes the autoregressive coupling crucial for accurate retrieval and ranking. Thus, we utilize the ranking loss 
$\mathcal{L}_{\text{rank}}$ to recover the retrieval/ranking ability. MTP is particularly useful for scenarios requiring dense, full-catalog scores with non-zero probability mass on every title.

\subsection{Efficient Training and Inference Stack}
\label{sec:method-stack}

The comparison above is enabled by a stack that makes large-scale fine-tuning and inference cheap. For \emph{training}, we combine four standard
components and report their integrated effect (Section~\ref{sec:e2e}): the
\textbf{Muon} optimizer \citep{jordan2024muon}, which orthogonalizes matrix
gradient updates and stores only momentum, roughly halving optimizer-state memory
and converging in fewer steps than our AdamW ~\cite{loshchilov2018decoupled} baseline; \textbf{on-the-fly sequence packing} with
block-diagonal masks \citep{krell2023efficient} plus \emph{cross-rank batch mixing}
that assigns different tasks to different data-parallel ranks in one global batch,
improving throughput and stability; \textbf{Cut Cross-Entropy}
\citep{wijmans2024cce}, a fused kernel that avoids materializing the
large-vocabulary logits tensor; and \textbf{HSDP} \citep{zhao2023fsdp}, which
shards state within a node and replicates across nodes, confining the heaviest
communication to high-bandwidth intra-node links ($\sim$7\% absolute MFU gain on
the 8B backbone). For \emph{inference} we serve on vLLM
(Section~\ref{sec:method-beam}). 

\section{Experiments}
\label{sec:experiments}

\subsection{Experimental Setup}
\label{sec:setup}

\paragraph{Confidentiality and training data.}
We study the streaming-media recommendation domain. By \emph{industry-scale} we
mean an 8B-parameter production backbone fine-tuned over a proprietary catalog on
$\geq 10M$ sampled examples \emph{per task} drawn from production logs, evaluated
under a production training- and inference-cost budget rather than on a small
retrieval-only academic benchmark. To respect confidentiality
we report \emph{relative} rather than absolute figures for proprietary quantities
(catalog size, corpus sizes, and dollar costs) and reported metrics (e.g., MRR, IFEval, MMLU). A retrieval task uses
member interaction history data ($\geq 10M$) snapshotted at a fixed date-hour; a content QA task uses an internal content QA corpus  ($\geq 10M$) derived from our rich content metadata;
an instruction-following task uses a mix of open-source datasets\citep{taori_alpaca_2023, lambert2024tulu3}  ($\geq 10M$). \sid{}s come from the
latest quantizer based on RQ-VAE~\cite{rajput2023tiger} snapshot.

\paragraph{Models and baseline.}
All the experiments are based on open-weight \emph{Qwen3-8B} backbone \citep{yang2025qwen3} with an
optional 0.6B variant. We fine-tune such pre-trained models by supervised fine-tuning (SFT) plus LoRA~\citep{hu2022lora} on our training data, and we conduct additional reinforcement learning (RL) after the SFT for language metric enhancement. We use up to 256 NVIDIA H200 and A100-80GB GPUs for our model training and inference.
The \emph{baseline} model in this paper refers to the previous \textbf{non-\sid{}} model: a LoRA-only recipe with a catalog-sized auxiliary head for ranking. The additional head was necessary for the baseline to generate logits across our entire catalog. 
However, this ranking head can degrade training efficiency and increase maximum memory usage.

\paragraph{Content QA corpus.}
Content QA answers natural-language questions about a title (e.g., ``What is
\emph{Movie A} about?''). We internally developed a large-scale
\emph{Content QA} corpus in which each title is referenced by
\emph{both} its text and its \sid{}, so the entity representation matches what
the model sees during retrieval. The corpus is generated synthetically by
prompting a frontier teacher model (e.g., \emph{GPT 5.2}) with internal title$+$\sid{} context and spans
four complementary question types: title-grounded descriptions, \sid{}-centric
semantics, member-like query grounding, and two-title comparisons. To reduce
sensitivity to prompt style, we sample system prompts at training time, including
a 20\% probability of an empty system message.

\paragraph{Evaluation.}
Ranking is measured with normalized Mean Reciprocal Rank (nMRR). For
queries $\mathcal{Q}$,
\begin{equation}
  \mathrm{MRR}=\frac{1}{|\mathcal{Q}|}\sum_{x\in\mathcal{Q}} \frac{1}{\mathrm{rank}_x},
  \qquad
  \mathrm{nMRR}=\frac{\mathrm{MRR}}{\mathrm{MRR}_{\text{ideal}}},
  \label{eq:nmrr}
\end{equation}
where $\mathrm{rank}_x$ is the position of the first relevant entity and the
normalizer accounts for the number of eligible relevant items per query. We
report the \textbf{average} ranking performance using an internal test set across 7 different offsets (i.e., predict $+1 - +7$ days future interactions).
We also offer detailed metrics computed only on discoveries (i.e., watching new content) and new users (e.g., users with $\leq K$ engagement) and items (e.g., titles launched after training cutoff), along with overall ranking.
Content QA is scored by an LLM judge (\emph{Qwen2.5-32B-Instruct})
\citep{zheng2023judge} on a stratified test set. Among numerous language benchmarks, we report 
IFEval \citep{zhou2023ifeval} and MMLU \citep{hendrycks2021mmlu} in this paper. 

\begin{table}[t]
  \centering
  \caption{\textbf{Single-head beam search vs.\ multi-token prediction.}
  Relative quality change of \textbf{the single-head over the MTP}; positive values favor the single-head. Metrics are grouped by task family.}
  \label{tab:abl-mtp-quality}
  \setlength{\tabcolsep}{7pt}
  \begin{tabular}{llcc}
    \toprule
    Task family & Metric & \textit{Qwen3-8B} & \textit{Qwen3-0.6B} \\
    \midrule
    \multirow{4}{*}{Ranking}
      & Overall ranking (nMRR) & \up{+2.0\%} & \up{+3.5\%} \\
      & Discovery (nMRR)       & \up{+7.3\%} & \up{+12.6\%} \\
      & New user (nMRR)        & \up{+2.1\%} & \up{+4.2\%} \\
      & New item (nMRR)        & \dn{-0.8\%} & \up{+6.8\%} \\
    \midrule
    Content Understanding
      & QA Accuracy               & \up{+2.0\%} & \up{+12.7\%} \\
    \midrule
    \multirow{2}{*}{Language}
      & IFEval (accuracy)      & \up{+3.3\%} & \up{+2.3\%} \\
      & MMLU (accuracy)        & \up{+5.7\%} & \up{+4.8\%} \\
    \bottomrule
  \end{tabular}
\end{table}

\begin{table}[t]
  \centering
  \caption{Single-head vs.\ MTP: training and inference system metrics. Relative quality change of \textbf{the single-head over the MTP}. Single-head trains
  cheaper; MTP only wins inference throughput.}
  \label{tab:abl-mtp-system}
  \setlength{\tabcolsep}{4.5pt}
  \begin{tabular}{lccccc}
    \toprule
    Backbone & Training Runtime $\downarrow$ & MFU $\uparrow$  & Tokens/sec $\uparrow$ & Dollar Cost $\downarrow$   & Inference/sec  $\uparrow$  \\
    \midrule
    \textit{Qwen3-8B}
      & \green{$0.37 \times$}           & \green{$1.15 \times$}  & \green{$1.18 \times$} & \green{$0.38 \times$} & \red{$0.57 \times$} \\ 
    \midrule
    \textit{Qwen3-0.6B}
      & \green{$0.50 \times$}  & \green{$1.21 \times$}      & \green{$1.20 \times$}     & \green{$0.97 \times$} & \red{$0.18 \times$}  \\
    \bottomrule
  \end{tabular}
\end{table}

\subsection{Core Study: Single-Head Beam Search vs.\ Multi-Token Prediction}
\label{sec:core}

This is our central result. We take the \emph{same} backbone,
\sid{} tokenizer, data mixture, and training budget per token and train the model in two
different ways: (1) single-head autoregressive training (Eq.~\ref{eq:loss}) and constrained beam search inference (Eq.~\ref{eq:beamscore})
 and (2) multi-token prediction with our auxiliary ranking
loss (Eq.~\ref{eq:mtp}), i.e., MTP in its strongest form. This isolates the
generation paradigm as the only variable, making the comparison apples-to-apples
at production scale.

\paragraph{Quality: single-head wins where it counts.}
Single-head beam search beats MTP on all ranking
and content-QA metrics (Table~\ref{tab:abl-mtp-quality}). On the \emph{Qwen 3-8B} model, it improves
discovery (\up{7.3\%}), new user recommendation (\up{2.1\%}), overall retrieval (\up{2.0\%}), content understanding (\up{2.0\%}), and general-language benchmarks (IFEval \up{3.3\%}, MMLU \up{5.7\%}). 
On the \emph{Qwen3-0.6B} model, the single-head advantage is uniform and larger including
\emph{gains} on discovery (\up{12.6\%}) and new item recommendations (\up{6.8\%}). 
Across both scales, we attribute the ranking/QA boost of the single-head to autoregressive
coupling: letting code $c_\ell$ condition on $c_{<\ell}$ lets the model correct
early mistakes, whereas MTP's parallel heads must commit to all codes from a
single hidden state under a conditional-independence assumption. 

\paragraph{Cost: single-head is far cheaper to train, but MTP is much faster at inference.}
The quality advantage does not come with a training-cost premium; in fact, the reverse
(Table~\ref{tab:abl-mtp-system}). For the 8B model, single-head trains $2.7\times$ faster than MTP with $2.6\times$ lower dollar cost, and higher model FLOPs utilization (MFU) \& token per second. The \emph{only} win for MTP is raw inference throughput
(\up{75\%}), consistent with
the efficiency motivation of the MTP literature
\citep{gloeckle2024mtp,hou2025rpg}. 
In our deployment, this throughput edge does
not offset the quality and training-cost gaps, so we \textbf{adopt single-head beam search as the default} and keep MTP only for applications that need dense full-catalog scores.

\paragraph{Reconciling with prior work.}
We confirm MTP's inference advantage but overturn the \emph{quality}
claim that parallel prediction matches autoregressive generation. Two features of
our regime explain the gap: a moderate \sid{} length ($L{=}4$), where
autoregressive decoding is cheap and its coupling most valuable (vs.\ the long
$L{\ge}64$ \sid{}s that make beam search prohibitive in RPG); and multi-task
training, which lets the single head share representations across retrieval,
content QA, and language.

\begin{table}[t]
  \centering
  \caption{\textbf{Overall metric comparison (\emph{Qwen3-8B}) of the single-head with \sid{} vs.\ the non-\sid{} baseline with an auxiliary ranking head.}
  Relative quality change of \textbf{the \sid{} model over the baseline}; positive
  values favor the \sid{} model. Metrics are grouped by task family. We observe enhanced ranking and content understanding, while language regression due to \sid{} vocab extension. We confirm the language regression can be recovered by RL post training (e.g., IFEval \up{+12.8\%}).}
  \label{tab:overall-quality}
  \setlength{\tabcolsep}{7pt}
  \begin{tabular}{llc}
    \toprule
    Task family & Metric & Single-head with \sid{} vs.\ Non-\sid{} baseline \\
    \midrule
    \multirow{4}{*}{Ranking}
      & Overall ranking (nMRR) & \up{+4.7\%} \\
      & Discovery (nMRR)       & \up{+5.3\%} \\
      & New user (nMRR)        & \up{+4.7\%} \\
      & New item (nMRR)        & \up{+77.3\%} \\
    \midrule
    Content Understanding
      & QA Accuracy               & \up{+21.7\%} \\
    \midrule
    \multirow{2}{*}{Language}
      & IFEval (accuracy)      & \dn{-7.3\%} \\
      & MMLU (accuracy)        & \dn{-10.0\%} \\
    \bottomrule
  \end{tabular}
\end{table}

\begin{table}[t]
  \centering
  \caption{Training system metrics between the single-head with \sid{} vs. the non-\sid{} baseline, where both models are based on \emph{Qwen3-8B}. Multipliers are relative to the non-\sid{} baseline.}
  \label{tab:system}
  \begin{tabular}{lcccc}
    \toprule
    Model & Runtime $\downarrow$ & MFU $\uparrow$ & Tokens/sec $\uparrow$ & Dollar Cost $\downarrow$ \\
    \midrule
    Single-head with \sid &\green{0.36$\times$} &  \green{4.75$\times$}
      & \green{4.0$\times$} & \green{0.27$\times$} \\
    \bottomrule
  \end{tabular}
\end{table}

\subsection{Single-head Beam Search with \sid{} vs.\ the non-\sid{} Baseline}
\label{sec:e2e}

Having established the generation paradigm, we report the end-to-end gains of the
full recipe (single-head $+$ native \sid{} $+$ new training/inference stack) over the
previous \emph{non-\sid{}} baseline.

\paragraph{Overall comparison.}
The single-head model with \sid{} outperforms the non-\sid{} baseline across ranking and content QA (Table~\ref{tab:overall-quality}). The headline gains are \up{77.3\%} on new item nMRR,
\up{21.7\%} on content-QA accuracy, and \up{$\sim$5\%} on overall ranking \& discovery. 
The significant new item nMRR boost is direct evidence that \sid{} prefix sharing lets the model recommend
titles it never saw engagement for.
We observe regression on language metriccs such as IFEval (\dn{7.3\%}) and MMLU (\dn{10.0\%}). The key reason is that expanding the vocabulary with $LK$  \sid{} tokens (assuming $L$-level codebook with $K$ tokens per level) and co-training on them perturbs the instruction-following distribution. To address this regression, we perform reinforcement learning (RL) with instruction-following datasets after the SFT stage, which yields \up{+12.8\%} improvement of the IFEval metric.

\paragraph{System efficiency.}
The training optimizations we introduced transform training economics (Table~\ref{tab:system}). Compared to the non-\sid{} baseline without any training optimizations (Section~\ref{sec:method-stack}),
single-head model trains $2.8\times$
faster, raises MFU $\sim4.75\times$ and tokens/sec $\sim4.0\times$, and cuts training cost by
73\%. These savings let us launch more model and ablation runs per unit budget.

\subsection{Ablation Studies}
\label{sec:ablations}

\begin{table}[t]
  \centering
  \caption{\textbf{LoRA vs.\ full fine-tuning on the single-head setup.}
  Relative quality change of \textbf{LoRA over full fine-tuning}; positive values favor  LoRA. Metrics are grouped by task family. We choose LoRA by default because of better ranking metrics, especially the cold-start item recommendation.}
  \label{tab:abl-fullft-quality}
  \setlength{\tabcolsep}{7pt}
  \begin{tabular}{llcc}
    \toprule
    Task family & Metric & \emph{Qwen3-8B} & \emph{Qwen3-0.6B} \\
    \midrule
    \multirow{4}{*}{Ranking}
      & Overall ranking (nMRR) & \up{+0.3\%} & \up{+0.1\%} \\
      & Discovery (nMRR)       & \up{+1.9\%} & $0.0\%$ \\
      & New user (nMRR)        & \up{+0.9\%} & $0.0\%$ \\
      & New item (nMRR)        & \up{+6.2\%} & \up{+4.5\%} \\
    \midrule
    Content Understanding
      & QA Accuracy               & $0.0\%$    & $0.0\%$ \\
    \midrule
    \multirow{2}{*}{Language}
      & IFEval (accuracy)      & \dn{-5.2\%} & \up{+3.3\%} \\
      & MMLU (accuracy)        & \dn{-6.1\%} & \up{+3.1\%} \\
    \bottomrule
  \end{tabular}
\end{table}

\begin{table}[t]
  \centering
  \caption{System metrics for the LoRA FT vs.\ full FT ablation. \textbf{Multipliers are relative to LoRA} of the same backbone. Full FT is  on-par-or-better than LoRA across all system metrics and model sizes, as it avoids the runtime memory overhead and kernel launch latencies introduced by LoRA.}
  \label{tab:abl-system}
  \begin{tabular}{llcccc}
    \toprule
    Backbone & Variant & Runtime $\downarrow$ & MFU $\uparrow$ & Tokens/sec $\uparrow$ & Dollar Cost $\downarrow$ \\
    \midrule
    \emph{Qwen3-8B}
      & Full FT & $1.0\times$     & \green{$1.13 \times$} & \green{$1.02 \times$}  & $1.0\times$ \\
    \midrule
    \emph{Qwen3-0.6B}
      & Full FT & \green{$0.87\times$} & \green{$1.12\times$} & \green{$1.17\times$} & \green{$0.9\times$} \\
    \bottomrule
  \end{tabular}
\end{table}

\paragraph{LoRA vs.\ full fine-tuning (FT).}
\label{sec:abl-lora}
We choose LoRA by default compared to full FT, as LoRA ranking metrics are on-par-or-better than full FT (Table~\ref{tab:abl-fullft-quality}), especially the cold-start item recommendation metric. 
According to Table~\ref{tab:abl-system}, full FT is on-par-or-better than LoRA because its standard dense matrix operations avoid the extra kernel launch overheads and lower arithmetic intensity introduced by LoRA. 
On the 8B model, LoRA shows a language-metric drop relative to full FT
because we apply LoRA to the output layer, whereas on the 0.6B model (where LoRA is not applied to the output layer) the effect reverses. 
We plan to revisit full
FT given its simplicity and system performance.

\begin{table}[t!]
  \centering
  \caption{\textbf{Optimizer and content-QA-data ablations against single-head
  LoRA training.} The columns report Muon relative to AdamW and all tasks
  relative to all tasks without content QA, respectively; positive values favor
  the first variant. Metrics are grouped by task family.}
  \label{tab:abl-data}
  \setlength{\tabcolsep}{6pt}
  \begin{tabular}{llcc}
    \toprule
    Task family & Metric & Muon vs.\ AdamW & All tasks vs.\ no Content QA \\
    \midrule
    \multirow{4}{*}{Ranking}
      & Overall ranking (nMRR) & \dn{-0.1\%} & \dn{-0.2\%} \\
      & Discovery (nMRR)       & \dn{-0.9\%} & $0.0\%$ \\
      & New user (nMRR)        & \up{+0.1\%} & \dn{-0.1\%} \\
      & New item (nMRR)        & \up{+5.2\%} & \up{+4.0\%} \\
    \midrule
    Content QA
      & Accuracy               & \up{+0.1\%} & \up{+211.4\%} \\
    \midrule
    \multirow{2}{*}{Language}
      & IFEval (accuracy)      & \dn{-1.9\%} & \dn{-3.2\%} \\
      & MMLU (accuracy)        & \dn{-0.6\%} & \dn{-1.8\%} \\
    \bottomrule
  \end{tabular}
\end{table}

\paragraph{Muon vs\ AdamW optimizer.}
Holding everything else fixed,  we compare two popular optimizers: Muon~\cite{jordan2024muon} and AdamW~\cite{loshchilov2018decoupled}. As shown in Table~\ref{tab:abl-data}, they reach similar quality, while Muon adds a small item cold-start gain, but Muon's system performance is clearly superior; e.g., $\sim 20\%$ higher MFU and throughput, lower memory consumption, and larger training batch size. Thus, we set Muon as
the default for both single-head and MTP training.

\paragraph{Contribution of content-QA data.}
Table~\ref{tab:abl-data} tests whether
content understanding helps ranking. Removing \emph{all} content-QA tasks leaves ranking nearly unchanged while collapsing content QA accuracy notably. We do not observe the synergy between ranking and content QA tasks in the current SFT-only recipe. This motivates moving content QA into a CPT stage followed by ranking-focused SFT, which is a direction for future iterations, and one consistent with the CPT+SFT of PLUM \citep{he2025plum}.

\paragraph{Other ablations.} We also found that introducing a separate weight for \sid{} tokens in \eqref{eq:loss} and having the auxiliary ranking loss for MTP~\eqref{eq:mtp} significantly improve ranking metrics. Note that we leave other non-critical ablations such as various \sid{} codebook shape (e.g., fan-in or fan-out), \sid{} level (e.g., large 2-level or small 8-level), or different task mixture ratios as future work.

\section{Conclusion}
\label{sec:conclusion}

Inside a unified multi-task LLM at industry scale, \emph{single-head
autoregressive \sid{} generation with constrained beam search beats multi-token
prediction} on most product-critical metrics (ranking and content QA) and trains at less than half the cost on \textit{Qwen3-8B} scale, while MTP leads only on inference throughput.
Even after we strengthen MTP with a new auxiliary ranking loss, the single-head wins.
Furthermore, the end-to-end recipe, from efficient training optimizations with \sid{} to vLLM-served
constrained catalog-aware beam search, more than halves training time and cost and lifts
cold-start (\up{+77\%}) and content-QA (\up{+22\%}) over the non-\sid{} baseline. 
Our findings have some limitations: (1) most experiments are based on offline training and test data, and online A/B test results might exhibit different trends, (2) the LLM could be overfit to our tasks and input format, and the output quality could be sub-optimal for diverse input format and tasks, and (3) missing continual pre-training (CPT) or mid-training stage that can warm-up the \sid{} token embeddings and reduce the training data volume for the SFT stage.
We will probe the crossover in \sid{} shape and length, catalog size, and task mixture at which it tips. We will also extend both single-head and MTP frameworks to multi-stage training such as CPT+SFT and CPT+SFT+RL. Finally, we will diversify our input prompt style and format and utilize a larger number of tasks during the SFT, to make our LLM robust.

\bibliographystyle{plain}
\bibliography{references}

\end{document}